\documentstyle[12pt]{article}

\begin{document}
\setcounter{page}{1} \pagestyle{plain}
\begin{center}
\Large{\bf \vspace{3cm}Field Equations and Radial Solutions in a Non-commutative
Spherically Symmetric Geometry }\\
\small{\bf \vspace{2cm} Aref Yazdani \footnote{a.yazdani@stu.umz.ac.ir}}\\
\vspace{0.5cm} {\it Department of Physics,
Faculty of Basic Sciences,\\
University of Mazandaran,\\
P. O. Box 47416-95447, Babolsar, IRAN}
\end{center}
\vspace{2cm}
\begin{abstract}
\vspace{0.5cm}We study a noncommutative theory of
gravity in the framework of torsional spacetime.
This theory is based on a Lagrangian obtained by applying
the technique of dimensional reduction of non-commutative
gauge theory, and that the yielded diffeomorphism invariant
field theory can be made equivalent to a teleparallel
formulation of gravity. Field equations are derived
in the framework of teleparallel gravity through Weitzenb\"{o}k geometry. We
solve these field equations by considering a mass that is
distributed spherically symmetric in a stationary static
spacetime in order to obtain a noncommutative line element.
This new line element interestingly reaffirms the
coherent state theory for a noncommutative Schwarzschild
black hole. For the first time, we derive the Newtonian
gravitational force equation in the commutative
relativity framework, and this result could provide
the possibility to investigate examples in various
topics in quantum and ordinary theories of gravity.
\end{abstract}
\vspace{2cm}
\section{Introduction}
Field equations of gravity and radial solutions have
been previously derived in noncommutative geometry
[1,2]. The generalization of quantum field theory by
noncommutativity based on coordinate coherent state
formalism also cures the short distance behavior of
point-like structures [3-7]. In this method, the particle
mass $M$ instead of being completely localized at a point,
is dispensed throughout a region of linear size $\sqrt{\theta}$,
substituting the position Dirac-delta function, describing
point-like structures, with a Gaussian function, describing
smeared structures. In other words, we assume the energy
density of a static, spherically symmetric, particle-like
gravitational source can not be a delta function distribution
and will be given by a Gaussian distribution of minimal
width $\sqrt{\theta}$ as follows:
\begin{equation}
\rho_{\theta}(r) = \frac{M}{(4\pi\theta)^{3/2}}\exp(-r^{2}/4\theta)
\end{equation}
Furthermore, noncommutative gauge theory appears
in string theory [8-13]: the boundary theory of an
open string is noncommutative when it ends on D-bran
with a constant B-field or an Abelian gauge field
(particularly see Ref. [8]). Therefore, closed
string theories are expected to remain commutative as
long as the background is geometric. Recent evidence
has found a connection between non-geometry and closed
string noncommutativity and even non-associativity
[14-16]; approaches using dual membrane theories [17]
and matrix models [18] arrive at the same conclusion.\\
The ordinary quantum field theory is unable to present
an exact description of exotic effects of the inherent
non-locality of interactions, so we need a model to
provide an effective description of many of the non-local
effects in string theory within a simpler setting [19].
The model leads to the gauge theories of gravitation through
an ordinary class of dimensional reductions of noncommutative
electrodynamics on flat space, which then can be made
equivalent to a formulation of teleparallel gravity,
macroscopically describing general relativity.
Moreover, this model is developed by the parallel theories
of gravitation, giving a clear understanding of Einstein's
principle of absolute parallelism. It is defined by a
non-trivial vierbein field and formed by a linear connection.
For carrying non-vanishing torsion, this connection is
known as Wietzenb\"{o}ck geometry on spacetime.\\
This model is given appropriately by a noncommutative
Lagrangian and introduced by authors in Ref. [2].
Admittedly, This Lagrangian and the relevant explanations
will be the basis of our next general calculations. In this
paper are going to use the Greek alphabet $(\mu, \nu ,
\rho, ... = 0,1,2,3)$ to denote indices related to
spacetime, and the first half of the Latin alphabet
$(a,b,c,... = 0,1,2,3)$ to denote indices related to
the tangent space. A Minkowski spacetime whose Lorentz
metric is assumed to have the form of $\eta_{ab} =
diag(-1,+1,+1,+1)$. The middle letters of the Latin
alphabet $(i,j,k,... = 1,2,3)$ will be reserved for space
indices. The noncommutative Lagrangian is expressed as
\begin{equation}
\dot{L_{Gr} }= \frac{\chi_{0}}{e^{2}} \det(h_{\sigma
}^{\sigma^{'}})\eta^{\mu \mu^{'}}\Big[\frac{1}{4}\eta^{\nu
\nu^{'}}\eta_{\lambda \lambda^{'}}
\dot{T}^{\ \ \lambda}_{\mu \nu}\dot{T}^{\ \ \lambda^{'}}
_{\mu^{'} \nu^{'}}- \dot{T}^{\ \ \nu}_{\mu \nu}\dot{T}^
{\ \ \nu^{'}}_{\mu^{'} \nu^{'}}+\frac{1}{2}\dot{T}^{\ \
\nu^{'}}_{\mu \nu} \dot{T}^{\ \ \nu}_{\mu^{'}{\nu}^{'}}
\Big]
\end{equation}
In the usual way, having a Lagrangian, which describes
gravitation based on noncommutative background, is like
those of gauge theories written in terms of contractions
of its field strength, here represented by torsion of
Weitzenb\"{o}ck connection. Its behavior
under a local change of $\Delta_{\mu}$ is the main
invariance property of the particular combination
torsion tensor fields. Here $e$ is Yang-Mills coupling
constant, noncommutative scale determines the Planck
length, and the Planck scale of n-dimensional spacetime
is given by
\begin{equation}
k = \sqrt{16 \pi G_{N}} = e\mid Pfaff(\Theta^{AB})
\mid^{1/2n}
\end{equation}
In mass dimension $2$ the weight constant $\chi_{0}$ is
\begin{equation}
\chi_{0}= \mid Pfaff(\Theta^{AB})\mid^{-1/n}
\end{equation}
In which the commutative limit, it reduces to gravitational
constant. therefore, $\Theta^{AB}$ is a noncommutative
parameter, defined as
\begin{equation}
\Theta^{AB} =
             \Bigg(
               \begin{array}{cc}
                 \theta^{\mu\nu} & \theta^{\mu b} \\
                 \theta^{\mu b} & \theta^{ab} \\
               \end{array}
\Bigg) \rightarrow \theta^{\mu\nu} = \theta^{ab} = 0
\end{equation}
By considering the calculation of superpotential and energy-momentum
current with respect to noncommutative gauge potential,
given by $B^{\ \mu}_{a} = |\det(\theta^{\mu^{'}a^{'}})|^
{1/2n}\hat{\theta}^{\nu\mu}\omega_{a\nu}$, the version
of noncommutative gravitational field equations are produced.
$\omega_{a\nu}$ are gauge fields corresponding to the gauging
of the translation group, i.e., replacing $R^{n}$ by the
Lie algebra $g$ of local gauge transformations with gauge
functions and its relation with the verbien field is expressed
as: $h_{a}^{\mu} = \delta_{a}^{\mu}-e\theta^{\nu\mu} \omega
_{a\nu}$ and $\delta_{a}^{\mu}$ has the perturbative effect
in the trivial holonomic tetrad fields of flat space.\\
It is important to note that by applying the "dimensional
reduction of gauge theories", noncommutative electrodynamics
gauge field; shown by the noncommutative Yang-Miles theory,
reduces to the gauge theories of gravitation , which
naturally yields Weitzenb\"{o}ck geometry on the spacetime.
Also, the induced diffeomorphism invariant field theory
can be made equivalent to a teleparallel formulation
of gravity macroscopically describing general relativity.
In section 2 we show that our Lagrangian can be made
equivalent with general relativity. In section 3 we
are going to derive the field equations by utilizing
various definitions of teleparallel gravity. By simplifying
and solving the field equations, we obtain the line element
in the spherically symmetric space-time in section 4.
We continue our discussion with investigations about
the limiting cases of our line element and horizons
of noncommutative Schwarzschild black hole in this
method. Finally we show how the Newtonian gravitational
force equation can be derived from our line element
in the commutative limit in section 5.
\section{Equivalence with General Relativity}
In order to continue our discussion to achieve to
noncommutative field equations,
we should show how our model can be
coupled with general relativity. With respect to
the given relation of
\begin{equation}
\dot{\Gamma}^{\rho}_{\ \mu \nu} = {\Gamma}^{\rho}_
{\ \mu \nu} + {\dot K}^{\rho}_{\ \mu \nu}
\end{equation}
for the vanishing curvature of the Weitzenb\"{o}ck connection,
we have
\begin{equation}
\dot{R}^{\rho}_{\ \theta \mu \nu} = R^{\rho}_
{\ \theta \mu \nu} + \dot{Q}^{\rho}_{\ \theta \mu \nu}
\equiv 0
\end{equation}
where
\begin{equation}
R^{\rho}_{\ \theta \mu \nu} = \partial_{\mu}
\Gamma^{\rho}_{\ \theta \nu} - \partial_{\nu}
\Gamma^{\rho}_{\ \theta \nu} + \Gamma^
{\rho}_{\ \sigma \mu}  \Gamma^{\ \ \sigma}_
{\theta \nu} -\Gamma^{\rho}_{\ \sigma \nu}
\Gamma^{\ \ \sigma}_{\theta \mu}
\end{equation}
is the curvature of the Levi-Civita connection,
The above equations show that, whereas in general
relativity torsion vanishes, in teleparallel gravity
it is curvature that vanishes. We rewrite the Eq.(7)
based on their components in order to find the scaler of
$\dot{R}^{\rho}_{\ \theta \mu \nu}$, therefore we have
\begin{equation}
{\dot Q}^{\rho}_{\ \theta\mu\nu} = \big(\partial_{\mu}
{\dot K}^{\ \ \rho}_{\theta\nu} - \partial_{\nu}
{\dot K}^{\ \ \rho}_{\theta\mu} + {\dot T}^{\ \
\rho}_{\sigma\mu}{\dot K}^{\ \ \sigma}_{\theta\nu}
\\ - {\dot \Gamma}^{\ \ \rho}_{\sigma\nu}{\dot K}^{\ \
\sigma}_{\theta\mu} - {\dot \Gamma}^{\ \ \sigma}_
{\theta\mu}{\dot K}^{\ \ \rho}_{\sigma\nu} +
{\dot \Gamma}^{\ \ \sigma}_{\theta\nu}{\dot
K}^{\ \ \rho}_{\sigma\mu} \big) + {\dot K}^{\ \
\rho}_{\sigma\nu}{\dot K}^{\ \ \sigma}_{\theta\mu}
\ - {\dot K}^{\ \ \rho}_{\sigma\mu}{\dot K}^{\ \
 \sigma}_{\theta\nu}
\end{equation}
that is the tensor written in terms of the Weitzenb\"{o}ck
connection only. Like the Riemanian curvature tensor,
it is a 2-form assuming values in the Lie algebra of
the Lorentz group (see Ref. [21]). By taking appropriate
contractions it is easy to show that
\begin{equation}
\dot{Q}^{\rho}_{\ \theta\mu\nu} = ({\dot D}_{\mu}{\dot k}^{\rho}
_{\ \theta\nu} - {\dot D}_{\nu}{\dot k}^{\rho}_{\ \theta\mu}) +
{\dot K}^{\rho}_{\ \sigma\nu}{\dot K}^{\ \ \sigma}_{\theta\mu}-
\dot{K}^{\rho}_{\ \sigma\mu}{\dot K}^{\ \ \sigma}_{\theta\nu}
\end{equation}
By considering the Eq.(20) and the following term
\begin{equation}
-R = \dot Q \equiv \frac{1}{2} \Lambda_{\rho}^{\ \theta}\dot{Q}^{\rho}_{\ \theta\mu\nu}
dx^{\mu}\wedge dx^{\nu}
\end{equation}
we achieve to the scalar version of Eq.(7),
\begin{equation}
R \equiv (\dot{K}^{\mu \nu \rho}\dot{K}_
{\rho \nu \mu}- \dot{K}^{\nu}_{\ \mu \rho}\dot{K}_{\nu}
^{\  \mu\rho})+ \frac{2}{h}\partial_\mu (h\dot{T}^{\nu \mu}
_{\ \ \nu}).
\end{equation}
The Lagrangian of Eq.(2) can be written in a simple
form of
\begin{equation}
\dot{L}= \frac{\chi_{0}}{e^{2}}\det(h_{\sigma
}^{\sigma^{'}})\bigg(\dot{K}^{\mu \nu \rho}\dot{K}_
{\rho \nu \mu}- \dot{K}^{\nu}_{\ \mu \rho}\dot{K}_{\nu}
^{\  \mu\rho}\bigg)
\end{equation}
with a combination of Eqs.(12) and (13), $\dot{L}$ takes the
following form
\begin{equation}
\dot{L} = \frac{\chi_{0}}{e^{2}}\det(h_{\sigma
}^{\sigma^{'}})\bigg(R- \frac{2}{h}(\partial_\mu (h\dot{T}^{\nu \mu}
_{\ \ \nu}))\bigg).
\end{equation}
By considering the Eqs.(3,4) $\dot{L}$ exchanges to
\begin{equation}
\dot{L} = L - \partial_{\mu} \big(\frac{h}{8\pi G}\dot{T}^{\nu \mu}
_{\ \ \nu}\big)
\end{equation}
up to a divergence at the commutative limit; therefore,
the Lagrangian of Eq.(2) $\dot{L}$ is equivalent to the
Lagrangian of general relativity as follows
\begin{equation}
\dot{L} = \frac{-1}{16\pi G}\sqrt{-g}R
\end{equation}
is the Einstein-Hilbert Lagrangian of general relativity.
However, this result could be extended with many further
terms, but this is enough to derive a valid field
equations.
\section{Noncommutative Field Equations}
In this section, we are going to present a reformulation of
teleparallel gravity, (is made equivalent to general
relativity). Due to the introduced noncommutative Lagrangian
(2), we are able to derive the field equations similarly
to the teleparallel method.
Weitzenb\"{o}ck geometric definitions and some well-known
concepts of general relativity [22-24] and teleparallel
gravity are required, (more explanations about these equations can be
found in Ref. [23],[25],[26]). In 4-dimension,
the noncommutative action integral is given by
\begin{equation}
S = \int \dot{L}_{Gr} d^{4}x
\end{equation}
Under an arbitrary variation $\delta h_{a}^{\ \mu}$ of the
tetrad field, the action variation is written in the following
form
\begin{equation}
\delta S = \int\Xi^{\ a}_{\mu}\delta h^{\mu}
_{\ a} h d^{4}x
\end{equation}
Where
\begin{equation}
h\Xi^{\ a}_{\mu} = \frac{\delta \dot{L}_{Gr}}{\delta B_{a}^
{\ \mu}}\equiv \frac{\delta \dot{L}_{Gr}}{\delta h_{a}^{\ \mu}} =
\frac{\partial \dot{L}_{Gr}}{\partial h_{a}^{\ \mu}}-\partial
_\lambda \frac{\partial \dot{L}_{Gr}}{\partial_{\lambda}\partial h^{\
\mu}_{a}}
\end{equation}
is the matter energy-momentum tensor. (More definitions
about this tensor can be found in Ref. [27]). Now, consider
first an infinitesimal Lorentz transformation as
\begin{equation}
\Lambda_{a}^{\ b}= \delta_{a}^{\ b}+\varepsilon_{a}^{\ b}
\end{equation}
With $\varepsilon_{a}^{\ b}= -\varepsilon^{b}_{\ a}$.
because of such transformation the tetrad should be changed as
\begin{equation}
\delta h_{a}^{\ \mu} = \varepsilon_{a}^{\ b} h_{b}^{\ \mu}
\end{equation}
The requirement of invariance of the action under local
Lorentz transformation therefore yields
\begin{equation}
\int\Xi_{a}^{\ b} \varepsilon_{\ a}^{b} h d^{4}x = 0
\end{equation}
Since $\varepsilon_{\ a}^{b}$ is antisymmetric, symmetric
of energy-momentum tensor yields some specific results
that can be seen in Ref. [22]. Consider spacetime
coordinates that are transformed as follows
\begin{equation}
x^{' \rho}= x^{\rho}+ \zeta^{\rho}
\end{equation}
Whereby, we retrieve the tetrad in the form of
\begin{equation}
\delta h_{a}^{\ \mu} \equiv h_{a}^{'\ \mu}(x)- h_{a}^{\
\mu}(x)= h_{\ a}^{\rho}\partial_{\rho}\zeta^{\mu} -
\zeta^{\rho}\partial_{\rho}h_{a}^{\ \mu}
\end{equation}
Substituting in Eq.(18), we have
\begin{equation}
\delta S = \int\Xi^{\ a}_{\mu}[h_{\rho}^{\
a}\partial_{\mu}\zeta^{\rho} - \zeta^{\rho}\partial_
{\rho}h_{a}^{\ \mu}] h d^{4}x
\end{equation}
or equivalently
\begin{equation}
\delta S = \int[\Xi^{\rho}_{\ c}\partial_
{\rho}\zeta^{c}+\zeta^{c}\Xi^{\rho}_{\ \mu}\partial_
{\rho}h_{c}^{\ \mu} - \zeta^{\rho}\partial_{\rho}
h_{a}^{\ \mu}] h d^{4}x
\end{equation}
Substituting the identity
\begin{equation}
\partial_{\rho}h_{a}^{\ \mu} = \dot{A}^{b}_{\ a\rho}
h_{b}^{\ \mu}- \dot{\Gamma}^{\mu}_{\ \lambda\rho}
h_{a}^{\ \lambda}
\end{equation}
where $\dot{A}$ is the spin connection in teleparallel gravity.
The important property of teleparallel gravity is
its spin connection is related only to the
inertial properties of the frame, not to gravitation.
In fact, it is possible to choose an appropriate
frame in which it vanishes everywhere.
We know the above formula vanishes by the Eq.(42),
(see also Ref. [28]), and making use of the symmetric
of the energy-momentum tensor, the action variation
assumes the form of
\begin{equation}
\delta S = \int \Xi^{\ \rho}_{c}\big
[\partial_{\rho}\zeta^{c}+(\dot{A}^{c}_{\ b\rho}
- \dot{K}^{c}_{\ b\rho})\zeta^{b} \big] h d^{4}x
\end{equation}
Integrating the first term by parts and neglecting
the surface term, the invariance of the action yields
\begin{equation}
\int\big[\partial_{\mu}(h\Xi^{\ \mu}_{a})-(\dot{A}^
{b}_{\ a\mu}- \dot{K}^{b}_{\ a\mu})(h\Xi^{\ \mu}_{b})
\big] \zeta^{a} h d^{4}x = 0
\end{equation}
From arbitrariness of $\zeta^{c}$, under the covariant
derivative $\ddot{D}_{\mu}$, it follows that
\begin{equation}
\ddot{D}_{\mu} h\Xi^{\ \mu}_{a}\equiv \partial_{\mu}
(h\Xi^{\ \mu}_{a})-(\dot{A}^{b}_{\ a\mu}- \dot{K}^{b}
_{\ a\mu})(h\Xi^{\ \mu}_{b})= 0
\end{equation}
By identity of
\begin{equation}
\partial_{\rho}h = h\dot{\Gamma}^{\nu}_{\ \nu \rho}\equiv
h\big(\dot{\Gamma}^{\nu}_{\ \rho\nu}-\dot{K}^{\ \
\nu}_{\rho\nu} \big)
\end{equation}
the above conservation law becomes
\begin{equation}
\partial_{\mu} \Xi^{\ \mu}_{a} + \big(\dot{\Gamma}^
{\mu}_{\ \rho\mu}-\dot{K}^{\mu}_{\ \rho\mu} \big)\Xi^{\
\rho}_{a}-(\dot{A}^{b}_{\ a\mu}- \dot{K}^{b}_{\ a\mu})
\Xi^{\ \mu}_{b} = 0
\end{equation}
In a purely spacetime form, it reads
\begin{equation}
\ddot{D}_{\mu} \Xi^{\ \mu}_{\lambda}\equiv \partial
_{\mu}\Xi^{\ \mu}_{\lambda} + \big(\dot{\Gamma}^
{\mu}_{\ \mu\rho}-\dot{K}^{\mu}_{\ \rho\mu} \big)
\Xi^{\ \rho}_{\lambda}-(\dot{\Gamma}^{\rho}_{\
\lambda\mu}- \dot{K}^{\rho}_{\ \lambda\mu})\Xi^
{\ \mu}_{\rho} = 0
\end{equation}
This is the conservation law of the source of
energy-momentum tensor. Variation with respect to the
noncommutative gauge potential $B^{\ \mu}_{a}$
yields the noncommutative teleparallel version of
the gravitational field equations
\begin{equation}
\partial_{\sigma}(h\dot{S}^{\ \mu\sigma}_{a})-k
h\dot{J}^{\ \mu}_{a} = k h\Xi^{\ \mu}_{a}
\end{equation}
where
\begin{equation}
h\dot{S}^{\ \mu \sigma}_{a} = hh^{\ \lambda}_{a}
\dot{S}^{\ \mu\sigma}_{\lambda}\equiv -k\frac{
\partial\dot{L}}{\partial(\partial_{\sigma}h^{a}
_{\ \mu})}
\end{equation}
which defines the superpotential, For the gauge
current we have
\begin{equation}
h\dot{J}^{\ \mu}_{a}= - \frac{\partial\dot{L}}{\partial
B^{a}_{\ \mu}}\equiv - \frac{\partial\dot{L}}{
\partial h^{a}_{\ \mu}}
\end{equation}
Note that the matter energy-momentum tensor which is
defined in this relation appears as the
source of torsion; similarly, the energy-momentum
tensor appears as the source of curvature in
general relativity. Our computation has led us
to the following results:
\begin{equation}
\dot{S}^{\ \mu\sigma}_{a} = 2\dot{T}^{\ \mu\sigma}
_{a} -\dot{T}^{\sigma\mu}_{\ \ a}- h^{\ \sigma}_{a}
\dot{T}^{\ \eta\mu}_{\eta} + h^{\ \mu}_{a} \dot{T}
^{\eta\sigma}_{\ \ \eta}
\end{equation}
and
\begin{equation}
\dot{J}^{\ \mu}_{a} = \frac{1}{k} h^{\ \lambda}_{a}
\dot{S}^{\nu \mu}_{\ \ c} \dot{T}^{c}_{\ \nu \lambda}
- \frac{h^{\ \mu}_{a}}{h} \dot{L}+ \frac{1}{k} \dot{A}
^{c}_{\ a\sigma} \dot{S}^{\mu \sigma}_{\ \ c}
\end{equation}
for noncommutative superpotential and gauge current.
The lagrangian “$\dot{L}$̇” appears again in our equations,
but notice that this term cross its coefficient yields
a term purely based on field strength. This simplified
expression maintains equivalence to general relativity.
We can observe that the gravitational field equations
depend on the torsion only. Finally the field equations
can be written as:
\begin{equation}
\partial_{\sigma} \Bigg(h \big( 2\dot{T}^{\ \mu\sigma}
_{a} -\dot{T}^{\sigma\mu}_{\ \ a}- h^{\ \sigma}_{a}
\dot{T}^{\ \eta\mu}_{\eta} + h^{\ \mu}_{a} \dot{T}
^{\eta\sigma}_{\ \ \eta}\big)\Bigg) - kh \Bigg(\frac
{1}{k}h^{\ \lambda}_{a} \dot{S}^{\nu \mu}_{\ \ c}
\dot{T}^{c}_{\ \nu \lambda} - \frac{h^{\ \mu}_{a}}{h}
\dot{L}+ \frac{1}{k}\dot{A}^{c}_{\ a\sigma} \dot{S}^
{\mu \sigma}_{\ \ c}\Bigg) = k h\Xi^{\ \mu}_{a}
\end{equation}
Where $k = \frac{\chi_{0}}{e^{2}}$ is a constant.
These field equations are similar to teleparallel
field equations. Although it would be distinguished
with different field strength $\dot{T}^{\ \mu\sigma}_{a}$
which is given by the covariant rotational of
noncommutative gauge potential of $B^{\ \mu}_{a}$.
By considering the following equations from the
teleparallel theory (see for instance,[20],[26],[28])
\begin{equation}
\dot{T}^{a}_{\ \mu\nu} = \partial_{\nu}h^{a}_{\ \mu}
- \partial_{\mu}h^{a}_{\ \nu} + \dot{A}^{a}_{\ e\nu}
h^{e}_{\ \mu} - \dot{A}^{a}_{\ e\mu}h^{e}_{\ \nu}
\end{equation}
\begin{equation}
\dot{\Gamma}^{\rho}_{\ \nu\mu} = h^{\rho}_{\ a}
\partial_{\mu} h^{a}_{\ \nu} + h^{\rho}_{\ a}
\dot{A}^{a}_{\ b\mu}h^{b}_{\ \nu}
\end{equation}
\begin{equation}
\partial_{\mu} h^{a}_{\ \nu} - \dot{\Gamma}^{\rho}
_{\ \nu\mu} h^{a}_{\ \rho} + \dot{A}^{a}_{\ b\mu}
h^{b}_{\ \nu} = 0
\end{equation}
and
\begin{equation}
\dot{T}^{\rho}_{\ \nu \mu} = \dot{\Gamma}^{\rho}
_{\ \mu \nu}-\dot{\Gamma}^{\rho}_{\ \nu \mu}
\end{equation}
The field equations take the exact following form
\begin{equation}
\frac{\partial}{\partial x^{\sigma}}
(\dot{\Gamma}^{\sigma}_{\ a\mu}-\dot{\Gamma}^
{\sigma}_{\ \mu a}) - \frac{\partial}{\partial
x^{\mu}}\dot{\Gamma}^{\lambda}_{\ a\lambda} +
\frac{\partial}{\partial x^{\lambda}}\dot{\Gamma}
^{\lambda}_{\ a\mu} - \dot{\Gamma}^{\eta}_{\
a\lambda}\dot{\Gamma}^{\lambda}_{\ \mu \eta} +
\dot{\Gamma}^{\eta}_{\ a\mu}\dot{\Gamma}^
{\lambda}_{\ \lambda\eta} = \frac{\chi_{0}}
{e^{2}} \rho(r) \frac{\partial}{\partial x^{a}}
\frac{\partial}{\partial x^{\mu}},
\end{equation}
which unlike the left hand side of Eq.(39), is written purely
based on noncommutative field strength,
the above field equation is written in terms of
Weitzenb\"{o}ck connection only. Regarding
to the equivalency between corresponding
Lagrangians and the above simplified field
equations and applying the Eq.(34), we have
therefore
\begin{equation}
R_{a\mu} -\frac{1}{2}h_{a\mu}R = k\Xi_{a\mu}
\end{equation}
as equivalent with Einstein's field equations.
Note that the equation (45) is not Einstein's field equations
but the teleparallel field equations
made equivalent to general relativity.
And equivalent model of teleparallel field equations
with general relativity expressed in references in detail,
(see for instance [26],[28]).
We continue our discussion to derive noncommutative
line element by solving these field equations.
\section{Noncommutative Line Element}
Teleparallel versions of the stationary, static, spherically,
axis-symmetric, and symmetric of the Schwarzschild solution
have been previously obtained [31],[32]. Within a
framework inspired by noncommutative geometry,
We solve the field equations for a distribution of
spherically symmetrically mass in a stationary static
spacetime, like the exterior solution of
Schwarzschild (see also Ref. [23]). Then it is
natural to assume that the line element is as follows
\begin{equation}
ds^{2} = -f(\tilde{r})dt^{2}+g(\tilde{r})dr^{2} +
h(\tilde{r})\tilde{r}^{2}(d\theta^{2}+\sin^{2}
\theta d\phi^{2}).
\end{equation}
With a new radial coordinate defined as $r =
\tilde{r}\sqrt{h(\tilde{r})}$, the line element becomes
\begin{equation}
ds^{2} = -A(r)dt^{2}+B(r)dr^{2} + r^{2}(d\theta^{2}
+\sin^{2}\theta d\phi^{2}).
\end{equation}
Usually one replaces the functions $A(r)$ and $B(r)$
with exponential functions to obtain somewhat simpler
expressions for the noncommutative tensor components.
Hence, we introduce the functions $\alpha(r)$ and $\beta(r)$
by $e^{2\alpha(r)} = A(r)$ and $e^{\beta(r)} = B(r)$ to get
\begin{equation}
ds^{2} = -e^{2\alpha}dt^{2}+e^{2\beta}dr^{2}
+ r^{2}(d\theta^{2}+\sin^{2}\theta d\phi^{2}).
\end{equation}
Tetrad components of the above metric takes the
following form:
\begin{equation}
h^{a}_{\ \mu} =
             \ \left[
             \begin{array}{cccc}
                -e^{2\alpha} & 0   & 0  & 0  \\
                      0 & e^{2\beta}\sin\theta\cos\phi & r\cos\theta\cos\phi & -r\cos\theta\sin\phi \\
                      0 & e^{2\beta}\sin\theta\sin\phi & r\cos\theta\sin\phi & r\sin\theta\cos\phi  \\
                      0 & e^{2\beta}\cos\theta         & -r\sin\theta        &        0             \\
               \end{array}\right]
\end{equation}
Weitzenb\"{o}ck connection $\dot{\Gamma}^{\rho}_{\ \mu\nu}$
has following expression:
\begin{equation}
\dot{\Gamma}^{\rho}_{\ \mu\nu} = h^{\ \rho}
_{a}\partial_{\nu}h^{a}_{\ \mu}
\end{equation}
Now, we can calculate the non-vanishing components
of Weitzenb\"{o}ck connection as follows:
$$\Gamma^{0}_{\ 01} = -2\alpha^{'}, \
\Gamma^{1}_{\ 11} = 2\beta^{'},\
\Gamma^{1}_{\ 22} = -re^{-\beta},$$
$$\ \Gamma^{1}_{\ 33} = -re^{-\beta}
\sin^{2}\theta,\ \Gamma^{2}_{\ 12} =
\frac{e^{\alpha}}{r} = \Gamma^{3}_{\ 13},$$
\begin{equation}
\ \Gamma^{2}_{\ 21} = \frac{1}{r} =
 \Gamma^{3}_{\ 31},\ \Gamma^{2}_{\ 33} =
-\sin\theta\cos\theta, \ \Gamma^{3}_{\ 23}
= \Gamma^{3}_{\ 32} = \cot\theta.
\end{equation}
By replacing these components in Eq. (44), the noncommutative
tensors of Eqs.(52-54) for the left-hand side of the
field equations will produce the following expression
\begin{equation}
N_{\hat{t}\hat{t}} = \frac{1}{r^{2}}\big
(-4e^{-2\beta}+ 1 - \psi_{\theta} \big)
-\frac{2}{r} \beta^{'}e^{-2\beta} = \frac{
\chi_{0}}{e^{2}} \rho(r) \delta_{\hat{t}\hat{t}}
\end{equation}
\begin{equation}
N_{\hat{r}\hat{r}} = \frac{1}{r^{2}}\big(2e
^{-2\beta}+1 - \psi_{\theta} \big)+\frac{2}{r}
\alpha^{'}e^{-2\beta} = \frac{\chi_{0}}{e^{2}}
\rho(r) \delta_{\hat{r}\hat{r}}
\end{equation}
\begin{equation}
N_{\hat{\theta}\hat{\theta}} = N_{\hat{\phi}
\hat{\phi}} = \frac{1}{r}e^{-2\beta}\big
(r\alpha^{''}+ r\alpha^{'2} - r\alpha^{'}
\beta^{'}+\alpha^{'}-\beta^{'} - 1 \big)+
\alpha^{'2}e^{-2\beta} = \frac{\chi_{0}}
{e^{2}} \rho(r) \delta_{\hat{\theta}\hat{\theta}}
= \frac{\chi_{0}}{e^{2}}\rho(r) \delta_
{\hat{\phi}\hat{\phi}}
\end{equation}
Adding equations (52) and (53) we get simply
\begin{equation}
\frac{1}{r^{2}}\big( \psi_{\theta}-e^{-2\beta}
+e^{-2\beta}(\alpha^{'}-\beta^{'}) +1
\big) = \frac{\chi_{0}}{e^{2}} \rho(r)
\end{equation}
where $\alpha(r) \neq \beta(r)$. It should also be
noted that by recalling Eq.(48), we can consider the
limiting case for our solution assuming $(\alpha^{'}-
\beta^{'})= k$, which $k$ is a constant, and by considering
the time-coordinate, we can shift this constant to an
arbitrary value. It is possible, therefore, without loss
of generality to choose $k = 0$. It does not contradict
with Eq.(48) to set $\alpha^{'} = \beta^{'}$. According
to this analysis, the equation $N_{\hat{t}\hat{t}} =
\frac{\chi_{0}}{e^{2}}\rho(r)\delta_{\hat{t}\hat{t}}$
can be written as
\begin{equation}
\frac{-1}{r}\frac{d}{dr}[r(e^{-2\beta} - \psi_
{\theta}-1 )] = \frac{\chi_{0}}{e^{2}}\rho(r)
\end{equation}
For a perfect fluid in thermodynamic equilibrium, the
stress-energy tensor takes on a particularly simple form
\begin{equation}
\Xi^{\mu\nu} = (\rho + P)u^{\mu}u^{\nu} + pg^{\mu\nu}
\end{equation}
where the pressure $P$ can be neglected due to the
distribution of mass and the gravitational effects;
consequently, only one term will remain in the above
formula as follows
\begin{equation}
\Xi^{a\mu} = \rho(r) \frac{dx^{a}}{dt} \frac{dx^{\mu}}{dt}
\end{equation}
or
\begin{equation}
\Xi^{a\mu} = \rho(r) \delta^{a\mu}
\end{equation}
Therefore, for spherically symmetric distribution
of mass that depends on r-coordinate, we can write
\begin{equation}
m(r) = \int^{r}_{0} 4\pi r^{2} \rho(r) dr
\end{equation}
Note that the $\rho(r)$ is defined by Eq.(1). Indeed,
we introduce the same energy density indicated in the
noncommutative perturbation theory [30]
\begin{equation}
m(r) = M_{\theta}(r) = \frac{2M}{\sqrt{\pi}}\gamma
(\frac{3}{2},\frac{r^{2}}{4\theta}).
\end{equation}
Eq.(56) can be integrated to find
\begin{equation}
e^{-2\beta} = 1 - \frac{\chi_{0}}{4\pi e^{2}}
\frac{m(r)}{r} + \psi_{\theta}
\end{equation}
Where $\psi_{\theta}$ is a function that carries the
tetrad field factor and will be defined later by Eqs.
(67,68). Now by considering
\begin{equation}
e^{-2\beta} = -\frac{1}{h_{11}} = h_{00}
\end{equation}
the noncommutative line element for a spherically
symmetric matter distribution is therefore
\begin{equation}
ds^{2} = -\big(1 - \frac{\chi_{0}}{4\pi e^{2}}
\frac{m(r)}{r} + \psi_{\theta} \big)dt^{2}
+ \big(1 - \frac{\chi_{0}}{4\pi e^{2}}
\frac{m(r)}{r} + \psi_{\theta} \big)^{-1}dr^{2}
+ r^{2} \big(d\theta^{2} + sin^{2}\theta d\phi^{2}\big).
\end{equation}
The constant field of $\frac{\chi_{0}}{e^{2}}$ in terms
of Eqs.(3),(4),(5) can be retrieved as
\begin{equation}
\frac{\chi_{0}}{e^{2}} = \frac{|\theta^{\mu b}|}
{16\pi G_{N}}
\end{equation}
where $G_{N}$ is the Newtonian constant, and $|\theta
^{\mu b}|$ is determined by $\theta^{\mu b}= \theta^{21}
= -\theta^{12}\equiv \theta$. Where $\theta$ is a real,
antisymmetric and constant tensor, therefore, the above
equation can be simplified to yield:
\begin{equation}
\frac{\chi_{0}}{e^{2}} = \frac{\theta}{16\pi G_{N}}
\end{equation}
New line element (64) in particular depends on $\psi_{\theta}$,
and naturally $\psi_{\theta}$ has its origin on the
quantum fluctuations of the noncommutative background
geometry and originally comes from the field equations.
The presented solution for our field equations produces
\emph{naturally} some additional terms in comparison with
the solution of noncommutative version of general relativity,
(\emph{naturally}, because it has some additional terms
in its components). These terms appear in the new line
element because $\psi_{\theta}$ relates to the noncommutative
torsional spacetime and algebraic properties in spherically
symmetric solution of the tetrad fields. We have therefore,
 $\psi_{\theta}$ in a simplified following equation
\begin{equation}
\psi_{\theta} = \varepsilon^{\hat{r}\hat{\theta}\hat{\phi}}
\varepsilon_{\hat{r}\hat{\theta}\hat{\phi}}h^{\hat{r}}
_{\ \hat{r}}e^{-\beta}
\end{equation}
Definition $\varepsilon^{\hat{r}\hat{\theta}\hat{\phi}}
\varepsilon_{\hat{r}\hat{\theta}\hat{\phi}} = \frac{-6}{h^{2}}$
is applied here. (see also Ref. [29]). According to this
definition and Eqs.(49) and (51), Through simplification,
we find the following form of $\psi_{\theta}$
\begin{equation}
\psi_{\theta} \cong  \sum_{k = 2n} \sum_{n=1}
(\frac{\chi_{0}}{4\pi e^{2}}\frac{m(r)}{r})^{k} - \sum_{k = 2n+1}
\sum_{n=1}(\frac{\chi_{0}}{4\pi e^{2}}\frac{m(r)}{r})^{k}.
\end{equation}
Note that $\psi_{\theta}$ is considered with the lower
bound of $\sum$. If we want to consider at least the
second order of $\theta$ (which is proposed by Ref. [7])
for $\psi_{\theta}$, then it is natural to assume $n = 1$.
Therefore, two states for our line element will be produced:
the imperfect state and the perfect state. Let us now
consider the perfect state. There is a proof for this
state in terms of some theorems in mathematics that
allows us to introduce our line-element as an appropriate
description for a noncommutative spacetime. Combination
of these theorems with regard to our results is given by:
(Following the Ref.[33])\\
\textbf{Theorem.} Let $L$ be a perfect field. Recall that
a polynomial $f(x)\in L[x]$ is called additive if
$f(x + y) = f(x) + f(y)$ identically. It is easy to see
that a polynomial is additive if and only if it is of the form
\begin{equation}
f(x) = 1 - a_{0}x + a_{1}^{2}x^{2}- ... \pm
a_{n}^{n}x^{n} = \sum_{n = 0}
a_{n}^{2n}x^{2n} -\sum_{n = 0} a_{n}^{2n+1}x^{2n+1}
\end{equation}
The set of additive polynomials forms a noncommutative
field in which $(f \circ g)(x) = f(g(x))$. This field
is generated by scalar multiplications $x\mapsto ax$
for $a\in L$ and $x_{i} \in f(x)$ does not commute with
the $x_{j} \epsilon f(x)$. Note that $a$ can be a constant
field and it has given as $\approx\frac{\chi_{0}}{4\pi e^{2}}$
here. (see [33] and references cited therein). It is clear
that components of $f(x)$ can be exactly replaced with
components of $h_{00}$.\\
Regarding to other investigations toward descriptions of
noncommutative spacetime, we should expand our discussion
into a comparison method with the other line elements
presented for noncommutative spacetime. Ref. [7] suggests
the following line element for noncommutative Schwarzschild
spacetime suggested
\begin{equation}
ds^{2} = -\bigg(1- \frac{4M}{r\sqrt{\pi}}\gamma(3/2,r^{2}
/4\theta)\bigg)dt^{2} + \bigg(1- \frac{4M}{r\sqrt{\pi}}
\gamma(3/2,r^{2}/4\theta)\bigg)^{-1}dr^{2} + r^{2}d\Omega^{2}
\end{equation}
where $d\Omega^{2} = d\theta^{2} + sin^{2}\theta d\phi^{2}$
and $\gamma(3/2,r^{2}/4\theta)$ is the lower incomplete
gamma function
\begin{equation}
\gamma(3/2,r^{2}/4\theta) \equiv \int^{r^{2}/4\theta}_{0}
dt \sqrt{t}e^{-t}
\end{equation}
We note that non-vanishing radial pressure is a consequence
of the quantum fluctuation of the spacetime manifold leading
to an inward gravitational pull and preventing the matter
collapsing into a point. According to the line element
(64), in a neighborhood of the origin at $r\leq \theta$,
the energy density distribution of a static symmetric
and noncommutative fuzzy spacetime is described by Eq.(1),
which replaces the Dirac $\delta$ distribution by a smeared
Gaussian profile. Meanwhile, in the imperfect state,
our line element can be made equivalent to the line element
of Eq.(70), and it is expected to happen when $\psi_{\theta}$
vanishes. Assuredly it is due to vanishing of the tetrad
components $h^{a}_{\ \mu}$ in Eq.(49) or even Wietzenb\"{o}ck
connections in Eq.(51). \emph{It means that in absence of
torsional spacetime}, the coordinate coherent state will
be produced in the noncommutative field theory. It is
completely reasonable since coherent state theory is derived
in the noncommutative framework of general relativity, and
the torsion is not defined in general relativity.
This equivalency is shown with the following relation
$$1- \frac{M}{2r\sqrt{\pi}}
\gamma(\frac{3}{2},\frac{r^{2}}{4\theta})$$
\begin{equation}
\cong g_{00}^{coherent \ state}
= 1- \frac{4M}{r\sqrt{\pi}}\gamma(\frac{3}{2},
\frac{r^{2}}{4\theta}).
\end{equation}
According to this proof, the solution of the presented
noncommutative field equations in the imperfect state of
itself results in the exact solution of noncommutative
general relativity field equations through coordinate
coherent state of our line element.\vspace{1cm}
\subsection{Schwarzschild Black hole, Horizons}
In this paper we have not extended our discussion
into black holes, but our introduced equations can
be the basis of a subject on noncommutative back holes.
Indeed the calculation of event horizons of a noncommutative
Schwarzschild black hole would be done by the horizon equation
$-h_{r_{H}} = h^{11}(r_{H}) = 0$. Answers to this equation
are illustrated by Figs.(1),(2). Fig.(1) shows the behavior
of $h_{00}$ versus the horizon radii when $\psi_{\theta}$
vanishes. It is clear that $\psi_{\theta}$ vanishing
approximately results in $g_{00}$ of Eq.(70), Fig.(2)
shows the behavior of $h_{00}$ at the same conditions
when we have $\psi_{\theta}$.\vspace{4cm}
\begin{figure}[htp]
\begin{center}
\includegraphics{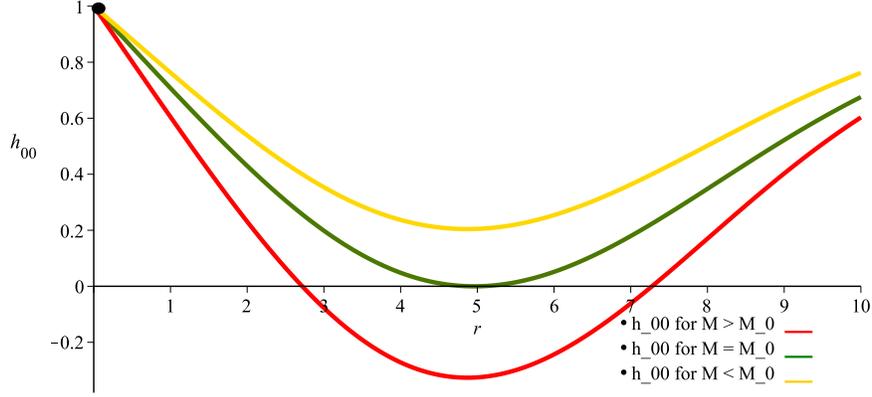}
\end{center}
\vspace{3cm}\caption{The imperfect state in a noncommutative
spherically symmetric geometry. The function of
$h_{00}$ vs $r\sqrt{\theta}$,
for various values of $M\sqrt{\theta}$, The upper curve
corresponds to $M = 1.00\sqrt{\theta}$ (without horizon),
the middle curve corresponds to $M = M_{0} \approx 1.90
\sqrt{\theta}$ (with one horizon at $r_{H} = r_{0} \approx
4.9\sqrt{\theta}$) and finally the lowest curve corresponds
to $M = 3.02\sqrt{\theta}$ (two horizons at $r_{H} = r_{-}
\approx 2.70\sqrt{\theta}$ and $r_{H} = r_{+} \approx 7.20
\sqrt{\theta}$).}
\end{figure}\vspace{3cm}
\begin{figure}[htp]
 \centering
  \includegraphics{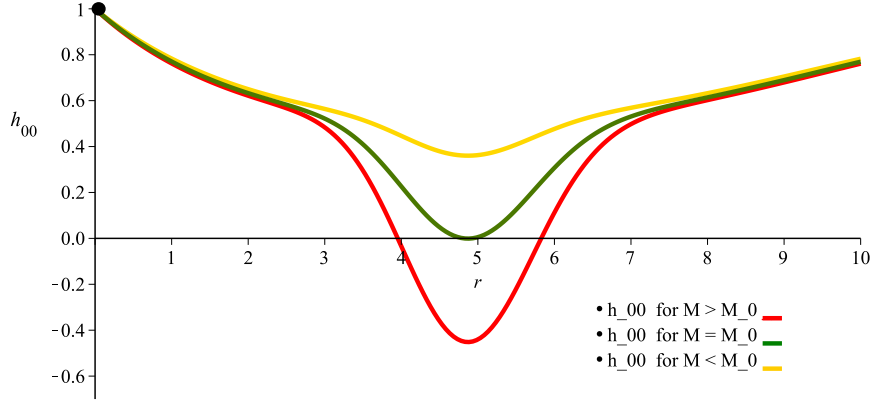}
\vspace{4cm}\caption{The perfect state in a noncommutative
spherically symmetric geometry. The function of $h_{00}$
vs $r\sqrt{\theta}$, for various values of $M\sqrt{\theta}$.
The upper curve corresponds to $M = 1.00\sqrt{\theta}$
(without horizon), the middle one corresponds to $M = M_{0}
\approx 1.90 \sqrt{\theta}$ (with one horizon at $r_{H} =
r_{0}\approx 4.9\sqrt{\theta}$) and finally the lowest curve
corresponds to $M = 3.02\sqrt{\theta}$ (two horizons at
$r_{H} = r_{-} \approx 3.90\sqrt{\theta}$ and $r_{H} =
r_{+} \approx 5.80 \sqrt{\theta}$).}
\end{figure}
\vspace{3cm}As we can see from these figures, there is
a different behavior in the perfect state in comparison
with the imperfect state near the horizon radii which is
due to the nature of torsional spacetime. Although, the
same behaviors have been indicated in the origin and
the higher bound of $r$.
\section{Force Equation in Commutative Limit}
In teleparallel gravity, the Newtonian force equation is
obtained by assuming the class of frames in which
the teleparallel spin connection $\dot{A}$ vanishes, and
the gravitational field is stationary and weak [29], [34].
In our model, the Newtonian gravitational force equation
directly derives from torsion components by $\psi_{\theta}$
in its commutative limit. When we write the expansion of
new line element in the noncommutative limit, we have:
\begin{equation}
h_{00} = 1- \frac{\chi_{0}}{4e^{2}}\frac{m(r)}
{r}+(\frac{\chi_{0}}{4e^{2}}\frac{m(r)}{r})^{2}-
(\frac{\chi_{0}}{4e^{2}}\frac{m(r)}{r})^{3}
\end{equation}
Or equivalent with
\begin{equation}
h_{00} = 1- A\frac{m(r)}{r}+B\frac{m(r)^{2}}{r^{2}}
-C\frac{m(r)^{3}}{r^{3}}
\end{equation}
Due to noncommutative effects, $r$ in the denominator
vanishes, but in the limit case, when it goes to
the commutative limit, it is modified to the commutative
$g_{00}$ of Schwarzschild solution in addition to a force
equation much similar to Newtonian gravitational force
equation. Note that the induced gravitational constant
of Eq.(3) vanishes in the commutative limit and agrees
with that found in [35] using the supergravity dual of
noncommutative Yang-Mills theory in four dimensions.
Newton was the first to consider in his Principia an
extended expression of his law of gravity including an
inverse-cube term of the form
\begin{equation}
F = G\frac{m_{1}m_{2}}{r^{2}}+B\frac{m_{1}m_{2}}{r^{3}}
,\  B \ is\  a\  constant.
\end{equation}
He attempts to explain the Moon's apsidal motion by above
relation. In the commutative limit our metric can be
defined in the form of:
\begin{equation}
h_{00}^{commutative} \cong 1- \frac{2M}{r}+ \frac{4M^{2}}
{r^{2}}-\frac{8M^{3}}{r^{3}}
\end{equation}
Where $m(r)$ is given by Eq. (61), and in the commutative
limit it has the form of
\begin{equation}
\lim_{\theta \longrightarrow 0}{m(r)} = 2M
\end{equation}
By considering the following terms in the Equations of (75),\\
$\ \ \ \bullet\ $ 	relativistic limits  $G = 1$,\\
$\ \ \ \bullet\ $ 	set the $m_{1} = m_{2} = 2M, B = -2M$,\\
therefore, for our line element we can set
\begin{equation}
h_{00}^{commutative} = (g_{00}^{commutative\  Schwarszchild\  solution} + F(r)^{Newton})
\end{equation}
As we can see from the Eq.(64) and (68), (expansion of
new line element) the $h_{00}$ has two parts:
torsional and non-torsional parts, the above relation
states that in the limit of commutativity, torsional
parts reduce to force equation of $F(r)$ and non-torsional
part yields the $g_{00}$ of commutative Schwarzschild
solution.\\
Einstein's theory of general relativity attributes
gravitation to curved spacetime instead of being due
to a force propagated between bodies. Energy and momentum
distort spacetime in their vicinity, and other particles
move in trajectories determined by the geometry of
spacetime. Therefore, descriptions of the motions
of light and mass are consistent with all available
observations. Meanwhile, according to general relativity's
definition the gravitational force is a fictitious force
due to the curvature of spacetime because the gravitational
acceleration of a body in free fall is due to its world
line being a geodesic of spacetime [36]. Whereas, through
a \emph{weak equivalence principle} assumed initially in
teleparallel gravity [37], our results are reasonable and
we can conclude that: \textbf{The presented solution in its
commutative limit attributes the gravitation to a force
propagated between bodies, and the curved spacetime, or
sum of torsion and curvature. This result is Similar to
Einstein-Cartan theory of gravity [38].} \\ Moreover, the
different behaviour of Schwarzschild black hole horizon,
which is absent in the previous method, is due to force
of heavy pulling from the black hole in terms of this
introduced force. As can be seen in fig.(2), the intensity
of this force has a direct relation with mass $M$, so the
heavier the black hole is, the stronger force it has near
its horizon.\\
\section{Conclusion}
In this letter, we have utilized a noncommutative
Lagrangian which gives us possibilities to use teleparallel
gravity to derive field equations. Solution of these field
equations in the spherically symmetric geometry yields a
new noncommutative line element. In the limit cases when
the torsion vanishes, we have obtained an interesting result:
\textbf{in absence of torsional spacetime the version of
coordinate coherent state in noncommutative field theory
will be produced}. Incidentally, Figs.(1),(2) show other
limit cases in our solution at the large distances and
different range of masses.\\
As we expressed before, there are conceptual differences,
in general relativity, curvature is used to geometrize the
gravitational interaction, geometry replaces the concept
of force, and the trajectories are determined, not by force
equations, but by geodesics. Teleparallel Gravity, on the
other hand, attributes gravitation to torsion. Torsion,
however, accounts for gravitation not by geometrizing the
interaction, but by acting as a force [29]. This is a
definition used in teleparallel gravity, whereas our model
do not exactly coincide with teleparalel gravity (in the
limit case only); therefore, it is natural to have more
complex results especially \textbf{the definition of existing
force in torsion for gravitational interactions is approved
clearly in the limit of commutativity in our model.}
Absolutely, attributing the gravitation to the force
equation in the relativity framework which is shown
directly through the commutative limit of our line element,
can be utilized in the various branches of physics.\\
\\
\\
\\
\\
\Large{\bf Acknowledgment}\\
\\
\small{I would like to give very special thanks to
Professor Kourosh Nozari for valuable comments on
this work. I also would like to thank Mr. Arya
Bandari for effective proofreading of this paper.}\\
\\
\\
\\


\begin{thebibliography}{7}
\bibitem{1}
M. R. Douglas and N. A. Nekrasov, \emph{Noncommutative
Field Theory}, Rev. Mod. Phys. \textbf{73} (2001), 977.\\
W. Kalau and M. Walze, \emph{Gravity, noncommutative
geometry and the Wodzicki residue}, J. Geo. Phys.
\textbf{16} 327 (1995), [arXiv:gr-qc/9312031].\\
K. Nozari and B. Fazlpour, \emph{Some consequences of
spacetime fuzziness}, Chaos Solitons Fractals \textbf{34}(2), (2007) 224-234.\\
P. Nicolini, \emph{Noncommutative black holes, the final
appeal to quantum gravity: a review}, Int. J. Mod. Phys.
A \textbf{24} (2009), 07 1229-1308.
\bibitem{2}
R. J. Szabo, \emph{Quantum field theory on noncommutative
spaces}, Phys. Rep. \textbf{378} (2003), (4) 207-299.\\
E. Langmann and R. J. Szabo, \emph{Teleparallel gravity
and dimensional reductions of noncommutative gauge theory},
Phys. Rev. D. \textbf{9} (2001), (10) 104019.
\bibitem{3}
X. Calmet and A. Kobakhidze, \emph{Noncommutative General
Relativity}, Phys. Rev. D \textbf{72} (2005), 045010
[arXiv:hep-th/0506157].\\
X. Calmet and A. Kobakhidze, \emph{Second Order Noncommutative
Corrections to Gravity}, Phys. Rev. D. \textbf{74} (2006),
047702 [arXiv:hep-th/0605275].\\
A. H. Chamseddine, \emph{Deforming Einstein's Gravity},
Phys. Lett. B. \textbf{504}(2001), 33 [arXiv:hep-th/0009153].\\
P. Aschieri et al, \emph{A Gravity Theory on Non-commutative
Spaces}, Class. Quant. Grav. \textbf{22} (2005), 3511
[arXiv:hep-th/0504183].
\bibitem{4}
A. Smailagic and E. Spallucci, \emph{Feynman path integral
on the noncommutative plane}, J. Phys. A. \textbf{36}
(2003), L467 [arXiv:hep-th/0307217].\\
K. Nozari and T. Azizi, \emph{Quantum mechanical coherent states
of the harmonic oscillator and the generalized uncertainty principle}
Int. J. Quant. Info. \textbf{3}(04), (2005) 623-632.\\
K. Nozari and S. Akhshabi,  \emph{Noncommutative geometry and
the stability of circular orbits in a central force potential},
Chaos Solitons Fractals, \textbf{37}(2), (2008) 324-331.
\bibitem{5}
A. Smailagic and E. Spallucci, \emph{UV divergence-free QFT on
noncommutative plane}, J. Phys. A. \textbf{36} (2003), L517
[arXiv:hep-th/0308193].
\bibitem{6}
A. Smailagic  and  E.  Spallucci,  \emph{Lorentz  invariance,
unitarity and  UV-finiteness of QFT  on noncommutative
spacetime}, J. Phys. A.  \textbf{37} (2004), 7169
[arXiv:hep-th/0406174].
\bibitem{7}
P. Nicolini, A. Smailagic and E. Spallucci, \emph{Noncommutative
geometry inspired Schwarzschild black hole}, Phys. Lett. B.
\textbf{632} (2006), 547 [arXiv:gr-qc/0510112].\\
K. Nozari and S. H. Mehdipour, \emph{Hawking radiation
as quantum tunneling from a noncommutative Schwarzschild
black hole} Class. Quant. Grav. \textbf{25}(17), (2008) 175015.
\bibitem{8}
C. S. Chu and P. M. Ho, \emph{Noncommutative open string and D-brane},
Nucl. Phys. B. \textbf{550} (1999), 151 [arXiv: hep-th/9812219].
\bibitem{9}
V. Schomerus \emph{D-branes and deformation quantization},
JHEP. \textbf{06} (1999), 030 [arXiv: hep-th/9903205].
\bibitem{10}
N. Seiberg and E. Witten, \emph{String theory and noncommutative
geometry}, JHEP. \textbf{09} (1999), 032 [arXiv: hep-th/9908142].
\bibitem{11}
F. Ardalan and H. Arfaei and M. Sheikh-Jabbari, \emph{Noncommutative
geometry from strings and branes}, JHEP.  \textbf{02} (1999),
016 [arXiv: hep-th/9810072].
\bibitem{12}
F. Ardalan and H. Arfaei and M. Sheikh-Jabbari, \emph{Dirac
quantization of open strings and noncommutativity in branes},
Nucl. Phys. B. \textbf{576 578} (2000), [arXiv: hep-th/9906161].
\bibitem{13}
C. Sochichiu, \emph{M[any] vacua of IIB}, JHEP  \textbf{0005}
(2000), 026 [arXiv: hep-th/0004062].
\bibitem{14}
R. Blumenhagen and E. Plauschinn, \emph{Nonassociative gravity
in string theory}, J. Phys. A. \textbf{44} (2011), 015401
[arXiv:1010.1263].
\bibitem{15}
D. L\"{u}st,  \emph{T-duality and closed string non-commutative
(doubled) geometry}, JHEP.  \textbf{12} (2010), 084 [arXiv:1010.1361].
\bibitem{16}
C. Condeescu and I. Florakis and D. L\"{u}st, \emph{Asymmetric orbifolds,
non-geometric fluxes and non-commutativity in closed string theory},
JHEP. \textbf{04} (2012), 121 [arXiv:1202.6366].
\bibitem{17}
C. S\"{a}mann and R. J. Szabo, \emph{Groupoids, loop spaces and
quantization of 2-plectic manifolds}, Rev. Math. Phys. \textbf{25} (2013),
1330005 [arXiv:1211.0395].
\bibitem{18}
A. Konechny A. Schwarz, \emph{Introduction to M(atrix) theory
and noncommutative geometry}, Phys. Rep. \textbf{360} (2002), (5) 353-465.\\
D. Mylonas and P. Schupp and R. J. Szabo, \emph{Membrane sigma-models and
quantization of non-geometric flux backgrounds}, JHEP. \textbf{09} (2012), 012
[arXiv:1207.0926].
\bibitem{19}
A. Chatzistavrakidis and L. Jonke, \emph{Matrix theory origins of
non-geometric fluxes}, JHEP. \textbf{02} (2013), 040 [arXiv:1207.6412].
\bibitem{20}
J. G. Pereira, \emph{Teleparallelism: A New Insight Into Gravity},
arXiv preprint. (2013), [arXiv:hep-th/1302.6983].
\bibitem{21}
P. Nicolini, \emph{Vacuum energy  momentum  tensor in (2+1) NC
scalar field theory}, [arXiv:hep-th/0401204].\\
A. Gruppuso, \emph{Newtons law in an effective non-commutative space-time},
J. Phys. A. \textbf{38} (2005), [arXiv:hep-th/0502144].
\bibitem{22}
S. Ansoldi and P. Nicolini and A. Smailagic and E. Spallucci,
\emph{Noncommutative geometry inspired charged black holes}, Phys. Lett.
B. \textbf{645} (2007), 261-266 [arXiv:0612035v1].\\
K. Nozari and S. Saghafi, \emph{Natural cutoffs and quantum
tunneling from black hole horizon}, JHEP \textbf{2012}(11), (2012) 1-18.\\
\bibitem{23}
D. McMahon, \emph{Relativity demystified. Tata McGraw-Hill
Education}, \textbf{103 104} (2006).
\bibitem{24}
{\o}. Gr{\o}n and S. Hervik, \emph{Einstein's general theory of
relativity: with modern applications in cosmology}, Springer Verlag. (2007).
\bibitem{25}
S. Weinberg, \emph{Gravitation and Cosmology} \textbf{126} Wiley, New
York, (1972).
\bibitem{26}
V. C. De Andrade and L. C. T. Guillen and J. G. Pereira,
\emph{Teleparallel gravity: an overview}, (2000), [arXiv:gr-qc/0011087].
\bibitem{27}
D. I. Olive, \emph{Lectures on Gauge Theories and Lie Algebras, University
of Virginia}, (1982), (notes taken by G. Bhattacharya and N. Turok).
\bibitem{28}
J. W. Maluf and M. V. O. Veiga and J. F. da Rocha-Neto,
\emph{Regularized expression for the gravitational energy-momentum
in teleparallel gravity and the principle of equivalence}, Gen.
Rel. Grav. \textbf{39} (2007), 3.\\
J. W. Maluf, \emph{The teleparallel equivalent of general relativity},
Annal. der. Phys. 525 (5) (2013), 339-357.
\bibitem{29}
R. Aldrovandi and J. G. Pereira, \emph{Teleparallel Gravity: An
Introduction } \textbf{(Vol. 173)} (2013), Springer.\\
V. C. de Andrade and J. G. Pereira, Phys. Rev. D. \textbf{56} (1997), 4689.
\bibitem{30}
K. Nozari and A. yazdani, \emph{The Energy Distribution of
a Noncommutative Reissner-Nordstr{\o}m Black Hole}. Chin.
Phys. Lett. \textbf{30} (2013), 9 090401.
\bibitem{31}
M. Sharif and M. J. Amir, \emph{Teleparallel Killing Vectors
of the Einstein Universe}, Mod. Phys. Lett. A. \textbf{23} (2008), 13.
\bibitem{32}
J. G Pereira and T. Vargas and C. M. Zhang, \emph{Axial-vector torsion and
the teleparallel Kerr spacetime}, Class. Quant. Grav. \textbf{18} (2001),
(5) 883.
\bibitem{33}
D. Goss, \emph{The Adjoint of the Carlitz Module and Fermat's s
Last Theorem. Finite fields and their applications},
\textbf{12} (1995), 165-188.\\
D. S. Thakur, \emph{Hypergeometric functions for function
fields. Finite Fields and Their Applications}, \textbf{12}
(1995), 219-231.\\
D. Goss, \emph{Basic structures of function field arithmetic},
(1997), Springer.\\
K. Kedlaya, \emph{The algebraic closure of the power series
field in positive characteristic}, Proc. Amer. Math. Soc.
\textbf{129} (2001), 3461-3470.\\
H. Hironaka, \emph{Resolution of singularities of an algebraic
variety over a field of characteristic zero: II}, Anna. Math.
(1964), 205-326.\\
B. Poonen, \emph{Fractional power series and pairings on Drinfeld
modules}, J. Amer. Math. Soc. \textbf{93} (1996), 783-812.
\bibitem{34}
G. Zet, Schwarzschild solution on a space-time with
torsion, (2003), [arXiv: gr-qc/0308078].\\
H. I. Arcos and J. G. Pereira, Torsion gravity: a
reappraisal, Int. J. Mod. Phys. D, 13 (2004), 2193-2240.
\bibitem{35}
N. Ishibashi and S. Iso and H. Kawai and Y. Kitazawa, "String scale
in noncommutative Yang-Mills," Nucl. Phys. B. 583, 159 (2000),
[hep-th/0004038].
\bibitem{36}
http://en.wikipedia.org/wiki/Newton's\ law\ of\ universal\ gravitation.
http://physicsessays.org/doi/abs/10.4006/1.3038751?journalCode=phes.
\bibitem{37}
G. J. Olmo, \emph{Violation of the equivalence principle
in modified theories of gravity}. Phys. rev. lett.
\textbf{98} (6) (2007), 061101.\\
R. Aldrovandi and J. G. Pereira and K. H. Vu,
\emph{Gravitation without the equivalence principle}, Gen.
Rel. Grav. \textbf{36} (1) (2004), 101-110.\\
C. Brans and R. H. Dicke, \emph{Mach's principle and
a relativistic theory of gravitation}. Phys. Rev.
\textbf{124} (3) (1961), 925.
\bibitem{38}
A. Trautman, \emph{Einstein-Cartan theory}, (2006), [arXiv:gr-qc/0606062].
\end{thebibliography}
 \end{document}